%
%
\magnification=\magstep1
\baselineskip=11pt plus .1pt minus .1pt
\hsize=12.5truecm
\vsize=19.0truecm  
\hfuzz=5pt\vfuzz=5pt
\tolerance=1000
\overfullrule=0pt
\parskip=0pt
\abovedisplayskip=3 mm plus6pt minus 4pt
\belowdisplayskip=3 mm plus6pt minus 4pt
\abovedisplayshortskip=0mm plus6pt minus 2pt
\belowdisplayshortskip=2 mm plus4pt minus 4pt
\predisplaypenalty=0
\clubpenalty=10000
\widowpenalty=10000
\parindent=2em
%
%
\font\pgnumfont=cmr9
\font\headlinefont=cmti9
 \font\titlefont=cmbx10
\font\authorfont=cmr10
\font\addressfont=cmti9
\font\datefont=cmr9
\font\sumfont=cmr9

\font\absfont=cmbx9
\font\secfont=cmr10
\font\subsecfont=cmti10
\font\subsubsecfont=cmr10
\font\figfont=cmr9
\font\figheadfont=cmbx9
\font\tabfont=cmr9
\font\tabheadfont=cmbx9
\font\mainfont=cmr10
\font\petitrm=cmr9

%
%
%
\newtoks\TITLE \newtoks\AUTHOR \newtoks\ADDRESS \newtoks\SUMMARY
\newdimen\sumindent \sumindent=\parindent
\newtoks\KEYWORDS \newtoks\SUBMITTED \newtoks\ACCEPTED
\newtoks\SENDOFF
%

%
%
\newtoks\firstpage
\let\firstpage=Y
\newtoks\AUTHORHEAD \newtoks\ARTHEAD \newtoks\VOLUME \newtoks\PAGES
\if!\the\AUTHORHEAD!\AUTHORHEAD={\the\AUTHOR}\fi
\if!\the\ARTHEAD!\ARTHEAD={\the\TITLE}\fi
\footline={\hfil}
\headline={\ifodd\pageno\rightheadline \else\leftheadline\fi}
\def\leftheadline{\if Y\firstpage\firsthead\global\let\firstpage=N
  \else\lefthead\fi}
\def\rightheadline{\if Y\firstpage\firsthead\global\let\firstpage=N
  \else\righthead\fi}
\def\lefthead{\pgnumfont\number\pageno\hfil\headlinefont\the\AUTHORHEAD}
\def\righthead{\headlinefont\the\ARTHEAD\hfil\pgnumfont\number\pageno}
\def\firsthead{\headlinefont Baltic Astronomy,~vol.\the\VOLUME,
\the\PAGES,~\the\year .\hfil}
\voffset=2\baselineskip 
%

\newdimen\oldbaselineskip \oldbaselineskip=\baselineskip
\def\test#1{\newlinechar=`@\if!\the#1! \message{#1 not given@}\fi}%
\def\printheader{
  \parindent=0pt
  \null\vskip1.cm
  \test{\TITLE}
  \vbox{\baselineskip=15pt
    \titlefont\the\TITLE
    }
  \vskip8mm plus8mm
  \test{\AUTHOR}
  \authorfont\the\AUTHOR
  \vskip2mm
  \test{\ADDRESS}
  \addressfont\the\ADDRESS
  \vskip2mm
  \test{\SUBMITTED}
  \line{\datefont Received \the\SUBMITTED
    \if!\the\ACCEPTED!\else, accepted \the\ACCEPTED\fi.\hfill}
  \vskip4mm plus4mm
  \vbox{\leftskip=\sumindent\parindent=0pt
    \parskip=5pt
    \absfont Abstract.
    \test{\SUMMARY}
    \sumfont\the\SUMMARY\par
    \absfont Key words:
    \test{\KEYWORDS}
    \sumfont\the\KEYWORDS\par
    }
  \sumfont
  \if!\the\SENDOFF!\else\footnote{}{Send offprint requests to:
 \the\SENDOFF}\fi
  \parindent=2em
  }
%
%
\newdimen\uppergap \newdimen\lowergap
\uppergap=5mm \lowergap=3mm
\newdimen\secind \newdimen\subsecind \newdimen\subsubsecind
\setbox0=\hbox{\secfont 9. }\secind=\wd0
\setbox0=\hbox{\subsecfont 9.9. }\subsecind=\wd0
\setbox0=\hbox{\subsubsecfont 9.9.9. }\subsubsecind=\wd0
\def\section#1{\goodbreak\par\vskip\uppergap
  \noindent\hangindent\secind\hangafter=1\secfont#1
  \vskip\lowergap\mainfont\par\nobreak}
\def\subsection#1{\goodbreak\par\vskip\uppergap
  \noindent\hangindent\subsecind\hangafter=1\subsecfont#1
  \vskip\lowergap\mainfont\par\nobreak}
\def\subsubsection#1{\goodbreak\par\vskip\uppergap
  \noindent\hangindent\subsubsecind\hangafter=1\subsubsecfont#1
  \vskip\lowergap\mainfont\par\nobreak}
%
%
\def\WFigure#1#2#3{\goodbreak\midinsert\vbox{
  \null\centerline{#2}\vskip1.5truemm
  \figheadfont\indent Fig.~#1.\figfont\ #3
  \par\mainfont
  }\endinsert}
%

%

%

%

%
\newdimen\tabind
\setbox0=\hbox{\tabheadfont Table 55.} \tabind=\wd0

%
%
\def\References{\vskip\uppergap
\line{\secfont REFERENCES\hfill}
  \vskip0.8\lowergap
 \petitrm
  }
\def\ref{\goodbreak
\hangindent12pt\hangafter=1
\noindent\ignorespaces}
\def\endref{\egroup}
%
%
\def\byebye{\egroup\par\vfill\supereject\end}
%
%

%
%

\def\utw{\smash{\rlap{\lower5pt\hbox{$\sim$}}}}
\def\udtw{\smash{\rlap{\lower6pt\hbox{$\approx$}}}}

\font\tbold=cmbx9
\font\tabfont=cmr9
\def\tablerule{\noalign{\vskip.9ex}\noalign{\hrule}\noalign{\vskip.7ex}}
\def\huad{\vrule height0pt depth0pt width5pt}  

\def\add{\vrule height0pt depth0pt width.5em}
\def\addu{\vrule height0pt depth0pt width1.3em}


\def\ddown{\lower2.5ex\hbox}
\def\ddow{\lower1.7ex\hbox}
\def\down{\lower1ex\hbox}
\def\uppp{\raise1ex\hbox}
\def\dnnn{\lower1ex\hbox}
\def\uuppp{\raise2ex\hbox}

\def\(o-c){$O-C$}


\def\angstr{A\kern-.56em\raise1.9ex\hbox{$\scriptscriptstyle\circ$}$\,$}

\newdimen\free\newdimen\shift
\def\Entry#1#2#3{\par\goodbreak\smallskip%
  \setbox1=\vbox{\advance\hsize by-10mm\parindent=0pt
    \def\\{\par}%
    \it#1. \rm#2}
  \line{\box1\hfill#3}\smallskip
}%
\newdimen\savesize

\def\shiftfigure #1#2#3#4#5{
    \vbox to #2 { \ifodd #5 \rightskip#4 \else\leftskip#4 \fi
                  \null\vfil
                  \figheadfont Fig.~#1.\figfont #3
                  \medskip
                }
                          }

\year1999

\input psfig.sty
{


\def\altaffilmark#1{{$^{#1}$}}
\def\altaffiltext#1#2{{\item{ }$^{\add #1}$ }{#2}}
\def\altwrap#1{{\item{}\addu #1}}
\def\up#1{\raise0.5ex\hbox{#1}}


\def\lsim{\hbox{$<\kern-1.7ex $\lower0.85ex\hbox{$\sim$} }}
\def\gsim{\hbox{$>\kern-1.7ex $\lower0.85ex\hbox{$\sim$} }}
\def\mean#1{{\langle}#1{\rangle}}   



\TITLE{DQ Herculis in Profile: Whole Earth Telescope Observations and
Smoothed Particle Hydrodynamics Simulations}

\VOLUME={8}
\PAGES={XXX--XXX}              
\pageno=1                      

\AUTHORHEAD={M. Wood and the XCov15 Team}

\AUTHOR{M. A. Wood\altaffilmark{1},
J. C. Simpson\altaffilmark{2},
S. D. Kawaler\altaffilmark{3},
M. S. O'Brien\altaffilmark{3},
R. E. Nather\altaffilmark{4},
T. S. Metcalfe\altaffilmark{4},
D. E. Winget\altaffilmark{4},
M. Montgomery\altaffilmark{4},
X. J. Jiang\altaffilmark{5},
E. M. Leibowitz\altaffilmark{6},
P. Ibbetson\altaffilmark{6},
D. O'Donoghue\altaffilmark{7},
J. Krzesinski\altaffilmark{8},
G. Pajdosz\altaffilmark{8},
S. Zola\altaffilmark{9,10},
G. Vauclair\altaffilmark{11},
N. Dolez\altaffilmark{11},
M. Chevreton\altaffilmark{12}
}

\ARTHEAD={DQ Herculis in Profile}

\ADDRESS={
\altaffiltext{1}{Department of Physics and Space Sciences
	and SARA Observatory,}
        \altwrap{Florida Institute of Technology,
	150 W. University Blvd., 
        Melbourne,}
	\altwrap{FL 32901-6988, wood@astro.fit.edu}
\altaffiltext{2}{Computer Sciences Raytheon, CSR 1310,
        P.O. Box 4127, Patrick Air}
   \altwrap{Force Base, FL 32925-0127 USA}
\altaffiltext{3}{Department of Physics, Iowa State University,
        Ames, IA 50011, }
   \altwrap{U.S.A.}
\altaffiltext{4}{Department of Astronomy and McDonald Observatory, 
        The University}
   \altwrap{of Texas, Austin, TX 78712-1083, U.S.A.}
\altaffiltext{5}{Beijing Astronomical Observatory, National
	Astronomical }
   \altwrap{Observatories, Chinese Academy of Sciences,
	Beijing, 100012, China }
\altaffiltext{6}{Department of Physics and Astronomy and Wise
        Observatory, Tel }
   \altwrap{Aviv University, Tel Aviv 69978, Israel}
\altaffiltext{7}{South African Astronomical Observatory, PO Box 9,
        Observatory}
   \altwrap{7935, South Africa}
\altaffiltext{8}{Copernicus Astronomical Center, ul. Bartycha 18,
        00-716 Warsaw, }
   \altwrap{Poland}
\altaffiltext{9}{Astronomical Observatory, Jagiellonian University,
	ul. Orla 171, }
   \altwrap{30-244 Cracow}
\altaffiltext{10}{Mt. Suhora Observatory, Pedagogical University,
	ul. Podchorazych }
   \altwrap{\ 2, 30-024 Cracow, Poland}
\altaffiltext{11}{Universit\'e Paul Sabatier, 
        Observatoire Midi-Pyr\'en\'ees, CNRS/}
   \altwrap{\ UMR5572,
        14 Ave. E. Belin, 31400 Toulouse, France}
\altaffiltext{12}{Observatoire de Paris-Meudon, DAEC, 92195 Meudon,
        France}
}

\SUMMARY={
DQ Herculis was the Summer 1997 WET northern-hemisphere primary
target.  The $O-C$ phase diagram of the 71-s signal reveals
out-of-eclipse phase variations resulting from the self
eclipses of a nearly edge-on, non-axisymmetric disc.
We work the forward problem of simulating DQ
Herculis numerically using the method of smoothed particle
hydrodynamics (SPH), and from the time-averaged structure calculate
the $O-C$ phase variations as a function of inclination angle.  The
WET and SPH $O-C$ phase variations match to a remarkable degree, and
suggest a system inclination in the range $89.5^\circ \le i \le 89.7^\circ$.  
}

\KEYWORDS={stars: binaries
cataclysmic variables
stars: individual: DQ Her
}

\SUBMITTED={September 25, 1999}

\vfill\eject

\printheader

\section{1. Introduction}

For relatively recent reviews of DQ Her systems, see Patterson (1994)
and chapter 9 of Warner (1995).  DQ Her was first studied as Nova 1934
Herculis, and Walker (1954, 1956) reported the then record-setting
orbital period of $\rm 4^h39^m$ and also the bewildering 71-s signal.
Warner et al.\ (1972) noted the phase of the this signal showed
significant and repeatable phase variations during eclipse ingress and
egress, but it was Bath, Evans, \& Pringle (1974) and Lamb (1974) that first
suggested the now-accepted magnetic rotator model for DQ Her,
identifying the source of the 71-s signal as the magnetic white dwarf
primary's spin period, and the pulse itself as reprocessed radiation
from the X-ray-emitting accretion column sweeping the inner accretion
disc (see Figure 1).

{\goodbreak\midinsert
\includegraphics{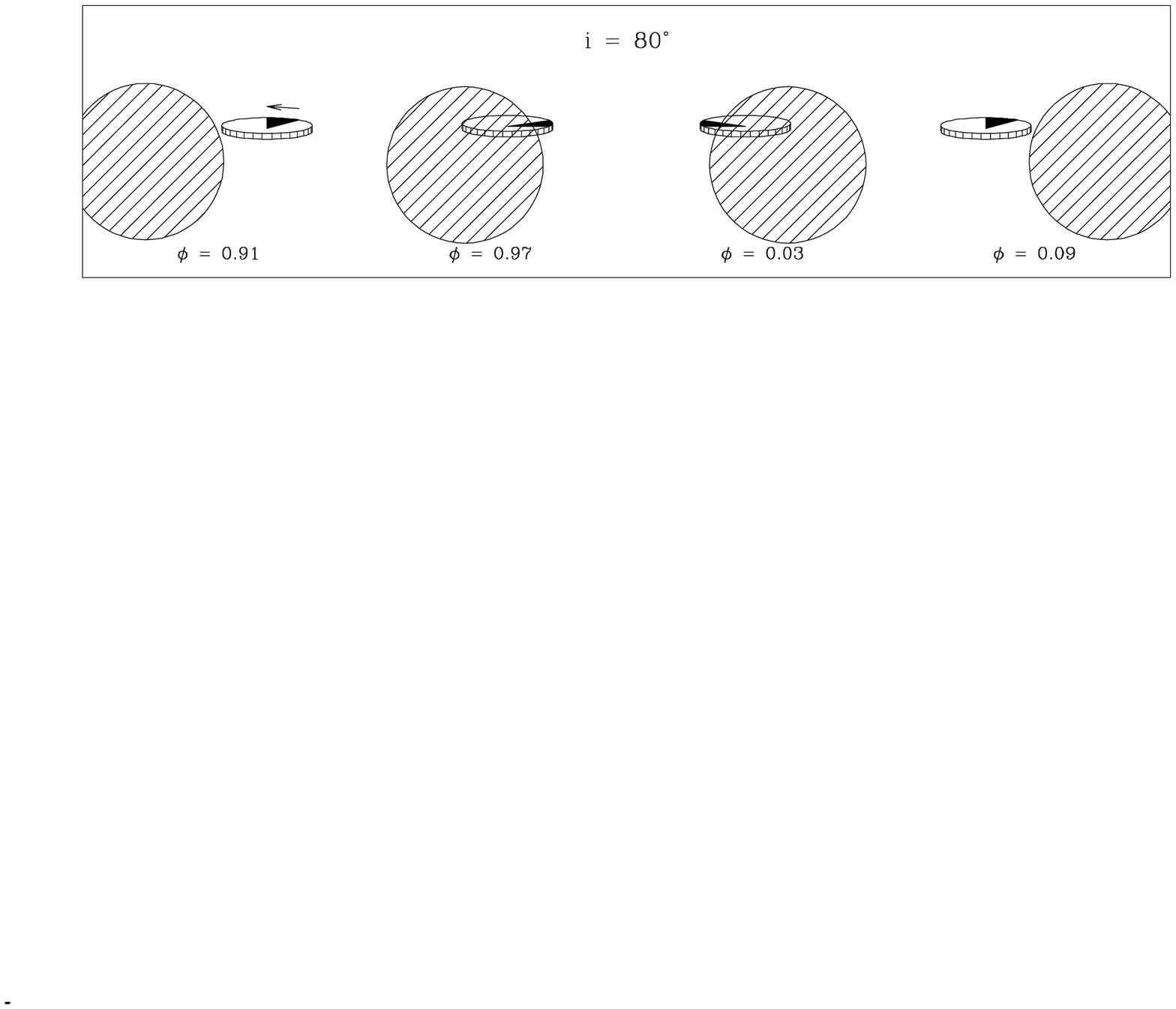}
\vbox{\vskip0.75in}
\WFigure{1}{}
{ A line drawing (approximately to scale) of the reprocessing
region of the disc in DQ Her passing through
eclipse, but viewed from an inclination angle of $i=80^\circ$.  The
EUV/X-ray beam is represented as a wedge on the disc surface, with the
direction of rotation indicated with the arrow. The accreting white
dwarf is located at the apex of the wedge.  
During ingress
(egress) the reprocessed pulses will arrive earlier (later) than when
out of eclipse.  Note that the non-zero disc opening angle results in
a front/back asymmetry in the visibility of the disc surface.  For
inclinations greater than $i=90^\circ-\mean{\theta}\approx86^\circ$,
the white dwarf is permanently eclipsed by the edge-on, optically-thick
accretion disc. }
\endinsert
}

Patterson, Robinson, \& Nather (1978; hereafter PRN) discovered
out-of-eclipse phase variations near orbital phase $\phi\sim0.7$, and
suggested this   was perhaps related to the expected thickening of the
disc downstream of the bright spot.  They also found that the pulse
arrival times were colour-independent, consistent with reprocessed
radiation.  Previous modeling attempts have tried to infer the disc rim
structure and system inclination by fitting to the observed $O-C$
phase diagram of PRN (see Chester 1979, Petterson 1980, O'Donoghue 1985).  

Here we work the {\it forward} problem of calculating the $O-C$ diagram from
the disc structure given by a 3-dimensional fluid dynamics simulation,
and then use these models to more tightly constrain the system
parameters for DQ Her.  The results presented here are a subset of
those detailed in Wood et al.\ (2000, in preparation).

\section{2. WET Observations}

The WET campaign included both large- and small-aperture telescopes,
but the results presented here only include data from the 2-m class
telescopes to avoid the necessity of a using a weighted average when
constructing an orbit-averaged light curve.  The Journal of
Observations for these runs is given in Table 1. 

\topinsert
$$\vbox{\tabfont
\tabskip=2em plus2em minus2em
\halign to 3in{
\hfil\huad #\hfil\huad&           
     \hfil # \hfil    &           
     \hfil # &           
     \hfil # \hfil\cr         
\multispan{4}{\tbold \hfil Table 1. Journal of Observations\hfil}\cr
\multispan{4}{\tbold  \hfil Large Aperture Telescopes\hfil}\cr
\tablerule
{Run} & {Start Time} & \omit \hfil{Length}\hfil &\ \  {Aperture}\cr
            & {(UTC)}     & \omit\hfil  {(s)}\hfil & {(m)}\cr
\tablerule
TSM-0023 & 1997 Jul \ 1 03:32:00 & 24,686 & 2.1 \cr
TSM-0024 & 1997 Jul \ 2 03:27:30 & 22,434 & 2.1 \cr
TSM-0025 & 1997 Jul \ 3 03:19:00 & 26,160 & 2.1 \cr
TSM-0027 & 1997 Jul \ 4 06:26:00 & 11,720 & 2.1 \cr
JXJ-0007 & 1997 Jul \ 4 12:56:59 & 16,608 & 2.2 \cr
TSM-0028 & 1997 Jul \ 5 03:42:00 &\ 9,498 & 2.1 \cr
JXJ-0009 & 1997 Jul \ 5 14:02:10 & 18,088 & 2.2 \cr
JXJ-0010 & 1997 Jul \ 7 12:46:50 & 16,104 & 2.2 \cr
GV-0521  & 1997 Jul \ 7 22:24:00 & 15,878 & 1.9 \cr
TSM-0030 & 1997 Jul \ 8 04:35:40 & 13,780 & 2.1 \cr
JXJ-0011 & 1997 Jul \ 8 12:43:10 &\ 9,208 & 2.2 \cr
JXJ-0012 & 1997 Jul \ 9 15:44:10 & 11,036 & 2.2 \cr
\tablerule
}}
$$
\endinsert

The data were reduced and barycentric calculations made using 
standard reduction techniques (Kepler 1990).  These were then 
converted to fractional
amplitude by dividing the original data by a copy which is
boxcar-smoothed with a window of 142 s in width, then subtracting
unity.  We use the method of Zhang et al.\ (1995) to obtain the
orbit-averaged light curve of the 71-s signal.  Specifically, we fold
our fractional amplitude data on the orbital period, but shift each
folded data segment by up to $\pm 35$ s to bring it into phase with
the mean \hbox{71-s} ephemeris.  We calculate the $O-C$ versus orbital phase
diagrams of the 71-s signal by fitting a 3-cycle sine curve to the
orbit-averaged lightcurve, shifting the window over the data by one
cycle between fits (see Figure 4, below).  

\vfill\eject
\section{3. SPH Simulations}

We have previously used our fluid dynamics code to explore the
3D hydrodynamics of superhumps (Simpson 1995; Wood \& Simpson 1995;
Simpson \& Wood 1998; Simpson, Wood, \& Burke 1998; and see Monaghan
1992 for a general review of SPH techniques). Here we use
the code to estimate the time-averaged equilibrium structure of the
disc in DQ Her.  For the runs presented here, we assume an ideal gas
law $P =(\gamma -1)\rho u$, where $u$ is the internal energy, and we
assume a ``nearly-isothermal'' $\gamma=1.01$.
We use the artificial viscosity prescription of Lattanzio et al.\
(1986) 
	$$
\Pi_{ij}=\cases{-\alpha\mu_{ij}+\beta{\mu_{ij}}^2
        & ${\bf v}_{ij}\cdot{\bf r}_{ij}\le 0$;\cr\cr{} \ \ \ \ \ 0 &
{\rm otherwise};\cr}
	$$
where
        $$
{\mu_{ij}}={{h{\bf v}_{ij}\cdot{\bf
r}_{ij}}\over {{c_s}_{,ij}(r_{ij}^2+\eta^2)}},
	$$
and ${c_s}_{,ij}={1\over 2}({c_s}_{,i}+{c_s}_{,j})$, ${\bf
v}_{ij}={\bf v}_i - {\bf v}_j$, and ${\bf r}_{ij}={\bf r}_i - {\bf
r}_j$.  We use $\alpha=0.5$, $\beta=0.5$, and $\eta=0.1h$ (but 
as we note below, we also tried $\alpha=1.5$, $\beta=1.0$, and the
results were nearly identical).  

The fundamental timestep for the system is set to be ${1\over
1000}P_{\rm orb}$, and individual particles can have shorter timesteps
if needed, in steps of factors of 2.  Because all particles have the
same size, an ancillary 3-D mesh of spacing $2h$ and the use of linked
lists allows vastly improved performance in the search/sort algorithm
to identify interacting neighbor particles.

For the system parameters, we use the results of Horne, Welsh, \& Wade
(1993), listed in Table 2.  

\topinsert
$$\vbox{\tabfont
\tabskip=2em plus2em minus2em
\halign to 2.5in{
\hfil\huad #\hfil\huad&           
     \huad\hfil # \hfil\cr         
\multispan{2}{\tbold \hfil Table 2. DQ Her System Parameters}\cr
\tablerule
{Parameter} & {Value} \cr
\tablerule
$P$ (hr)        & $4.56$ \cr
$q$             & $0.66$ \cr
$M_1/M_\odot$   & $0.60\pm 0.07$\cr
$M_2/M_\odot$   & $0.40\pm 0.05$\cr
$a/R_\odot$     & $1.41\pm 0.05$\cr
\tablerule
}}
$$
\endinsert

We start with an initially-empty disc  and begin by adding 2,000
particles per orbit for the first 10 orbits, after which time we hold
the total number of particles in the simulation at $N=20,000$, replacing
particles accreted onto the primary or secondary or lost from the
system.  In about 120 orbits, the system reaches a state of dynamical
equilibrium that should most closely approximate the structure of the
disc in DQ Her itself, and we follow the system for an additional 100
orbits for a total of 220 orbits calculated.  

Because the spatial resolution of the simulation is of order the SPH
smoothing length $h\approx0.005 a$ and the particle number density is
$\rho\sim 20-50 h^{-3}$, phase-space is coarsely binned and the
numerical statistical signal-to-noise ratio is low.  We circumvent
this problem by using the old numerical trick of averaging in time.
Sampling five times per orbit insures that consecutive samples are not
strongly correlated, and by concatenating the phase-space snapshots over
50 orbits in the co-rotating frame, we generate a 5,000,000-particle
{\it ensemble disc}.  We actually use consecutive 50 orbit spans from
the 100 orbits calculated following the establishment of equilibrium.
As expected, the two ensemble discs give effectively identical
results.  In addition, the solutions from 
higher viscosity runs with differing numbers of simulation particles
($N=8,000$ and 20,000, $\alpha=1.5$, $\beta=1.0$) also give essentially
identical results.

We calculate the opening angle $\theta(\psi)$ of the disc as a
function of the azimuth angle (in the co-rotating frame) $\psi$ by
binning the particles in the ensemble disc (minus the accretion
stream) into 120 $3^\circ$ sectors ($\sim$40,000 particles per
sector).  We then sort the particles in each sector by
$\theta=\tan^{-1}(|z|/r)$.  We make no attempt to calculate radiative
transfer in our calculations, and so our disc is too hot and has no
identifiable photosphere.  Instead, we assume that the photospheric
depth is proportional to the fractional depth in the sorted list.  To
give an example using round numbers, a given sector might have 10,000
particles in it, and we would calculate the ``95-th percentile
profile'' [$\equiv\theta_{95}(\psi)$]
as the angle of the 9,500-th particle, and the radius of the
reprocessing region in this sector is taken to be the mean radius of
the 200 particles with indices from 9,400 to 9,600.

\WFigure{2}{\psfig{figure=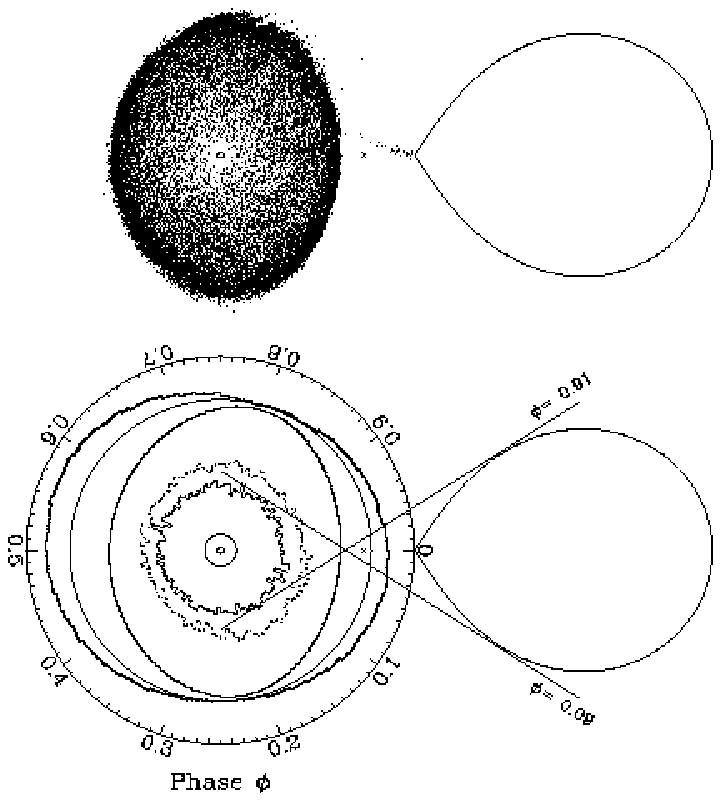,width=11cm,rheight=11.5cm,angle=0}}
{The $(x,y)$ disc projection.  In the top panel,
20,000 particles are shown and the system center of mass is indicated
with an ``$\times$''.  The bottom panel shows several features, as
described in the text.
}

The results of this analysis are shown in Figure 2.  The top panel
shows the zero-inclination view, including the positions of 20,000
particles.  The outer disc is non-circular and truncated beyond
$r\sim0.4a$ (see Paczy\'nski 1977; Osaki, Hirose \& Ichikawa 1993).
The bottom panel is a line drawing with several components.  First,
the orbital phase is labeled on a scale centered on the primary white
dwarf.  The secondary and white dwarf primary (innermost circle) are
both shown to scale.  The next circle out from the primary shows the
radius that has a 71-s Keplerian orbital period, which should be the
approximate Alfv\'en radius of the magnetic white dwarf.  The solid
polar histogram shows the radius of the $\theta_{99}$ profile and the
dotted one shows the $\theta_{95}$ profile in each sector -- this
should be roughly the radius of the reprocessing region $r_{\rm rp}$
as a function of $\psi$.  Note that both profiles are circular to a
good approximation and suggest $r_{\rm rp}\lsim0.2a$, or about
one-half of the disc radius and consistent with the radius given by
the eclipse-related phase shift which operates between orbital phases
$\phi\approx0.9$ and 0.1 (see e.g., Zhang et al.\ 1995, their Figure
9).  The match between the radius of the reprocessing region based on
the observations and that derived from the SPH calculations gives some
confidence in our approach.

The next solid line is simply the outline of the disc's outer radius --
specifically, the 99-th percentile radius profile. 

Between the outer radius of the disk and the scale, the Figure shows
the opening angle as a function of azimuthal angle as the heavy line
between the reference circle just beyond the 
disc's radial edge and the outer scale.  The radial scale
is such that $r=0$ corresponds to $0^\circ$, the reference circle
inside the scale corresponds to $2.73^\circ$, and the radius of the
scale corresponds to $3.51^\circ$.  The mean opening angle of the
95-th percentile profile is $2.94^\circ$, larger than the canonical
$h/r\sim0.015$ ($\theta\sim1^\circ$) suggested by the $\alpha$-model,
where $h$ is the disk semi-thickness.  Note however that the $\alpha$
model also predicts $h\propto r^{9/8}$ which would suggest that the
entire disk surface would be able to reprocesses radiation from the
X-ray beam, in contrast to both observations and the hydrodynamical
results here.

The opening angle of the reprocessing region is largest
where the outer radius of the disk is smallest, and vice versa.  If we
assume a constant density fluid and roughly Keplerian velocities, 
then the azimuthal mass flux is
proportional to the azimuthal velocity $v_{\psi}\propto r^{-1/2}$
times the cross-sectional area of the disk.  Therefore, if the outer
radius of a constant-height disk is decreased, the cross-sectional
area drops proportionately, but the velocity only increases as the
inverse square root.  Thus to satisfy the constraint that the
azimuthal mass flux is constant, the disk thickness must be largest in
quadrants with the smallest radial extent.  Looking back at Figure~2,
we see that indeed the surface density is highest where the outer disk
radius is smallest.  In large part, the rim profile of the
reprocessing region is the result of the non-axisymmetric shape of the
streamlines in the outer disk.

\WFigure{3}{\psfig{figure=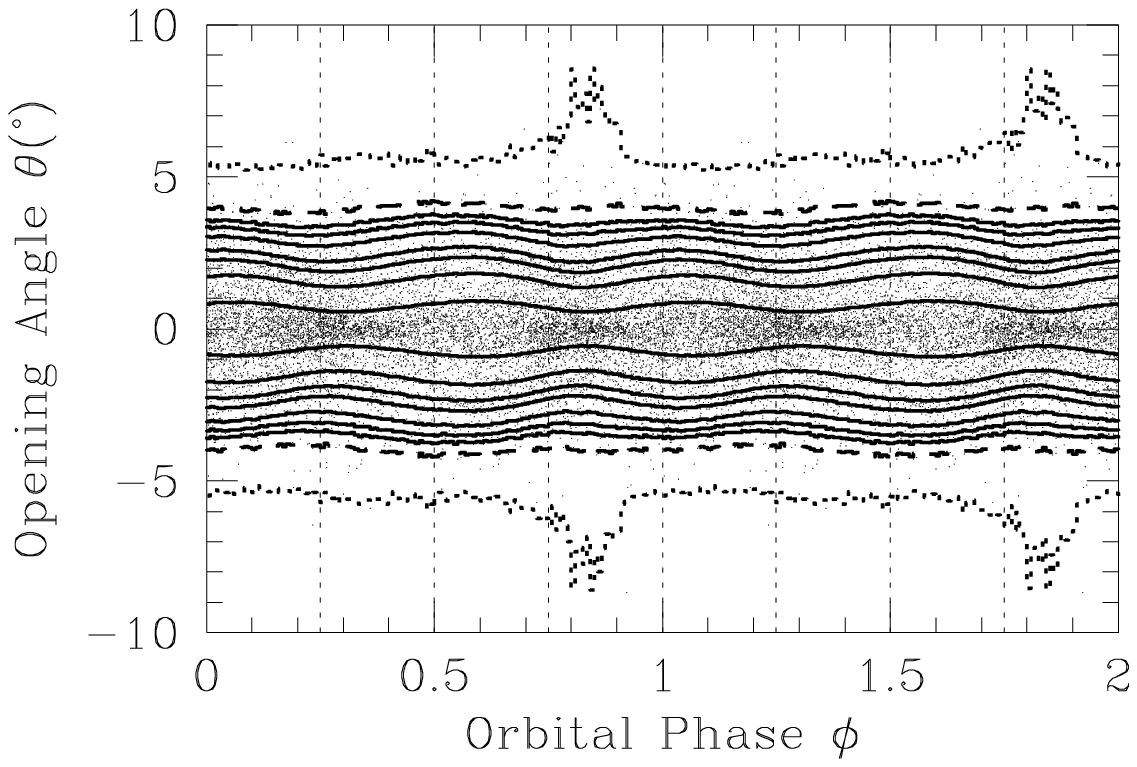,rwidth=11cm,width=16cm,rheight=6.5cm,angle=0}}
{The $(\phi,\theta)$ projection of the disc.   Only 30,000 particles
are shown.  The contour levels shown as solid lines are --- moving out from the
mid-plane --- 
$\theta_{50}$,
$\theta_{75}$,
$\theta_{85}$,
$\theta_{90}$,
$\theta_{95}$,
$\theta_{97}$, and
$\theta_{98}$.
The 
$\theta_{99}$ contour is shown as a dashed line, and the 
$\theta_{99.9}$ contour as a dotted line.
The thickening from the accretion stream is most
evident in the 99.9\% contour, but only marginally evident in the 99\%
contour.
}

Figure 3 shows the $\theta$ versus $\phi$ projection on the disk, with
the 50\up{th} through the 99.9\up{th} percentile contour levels
plotted over 30,000 points from the ensemble disc.  The
contours are very similar to each other over the range $\theta_{75}$
to $\theta_{98}$.  The result of this is that the calculated $O-C$
phase variations are not a sensitive function of which profile is used
over this range.  It also seems likely that our profiles would not
differ much if we'd included detailed radiative transfer to determine
the location of the disc photosphere.  Note that the accretion stream
doesn't significantly affect the shape of the profile until the
$\theta_{99}$ profile, and is only obvious in the $\theta_{99.9}$
profile.  These profiles give $O-C$ phase variations completely unlike
those observed, and so we conclude that the accretion stream must be
optically thin.

We integrate over the projected reprocessing region as detailed in
Wood et al.\ (2000, in preparation), including the light travel time
across the orbit and disc.  Weighting by azimuthal angle, we can
calculate the $O-C$ versus orbital phase diagram as a function of
inclination angle.  The results for the $\theta_{95}$ profile are
shown in Figure 4, where the calculated $O-C$ diagrams for inclination
angles ranging from 85 to $89.9^\circ$ are shown over the
orbit-averaged WET $O-C$ points.  The curves shown are for the 95-th
percentile profile, and the best match with observations are for
inclination angles between $89.5^\circ$ and $89.7^\circ$. 
Because it is $i-{1\over2}\theta$ to which these results are sensitive,  
if our calculated disk thickness is too large, then the inclination
of DQ Her is even closer to $90^\circ$.

The total extent of the $O-C$ phase variations is less than $0.5$
cycles, suggesting that we see reprocessed radiation from only one
magnetic pole and that the orbital period is 71 s and not 142 s for a
two-pole magnetic accretor (see Zhang et al.\ 1995).  However, the
observed points range up to 0.30 cycles at the beginning of eclipse
egress, and the  points appear to turn down at the end of eclipse
ingress.  Our current model cannot explain this observation, but we
speculate that the pattern of the source of the 71-s pulsed radiation
on the surface of the disk might show trailing ``arms'' beyond
the reprocessing region itself resulting from Keplerian shear of a
radial thermal pulse.  Such a pattern would give rise to the observed
phase behavior near totality of the reprocessing region.  It may also
be simply the result of a white dwarf rotation axis non-aligned with
the orbital axis or perhaps reprocessing off of vertical structure
beyond the calculated reprocessing region.  We are currently
attempting to resolve this subtle but important question.

\WFigure{4}{\psfig{figure=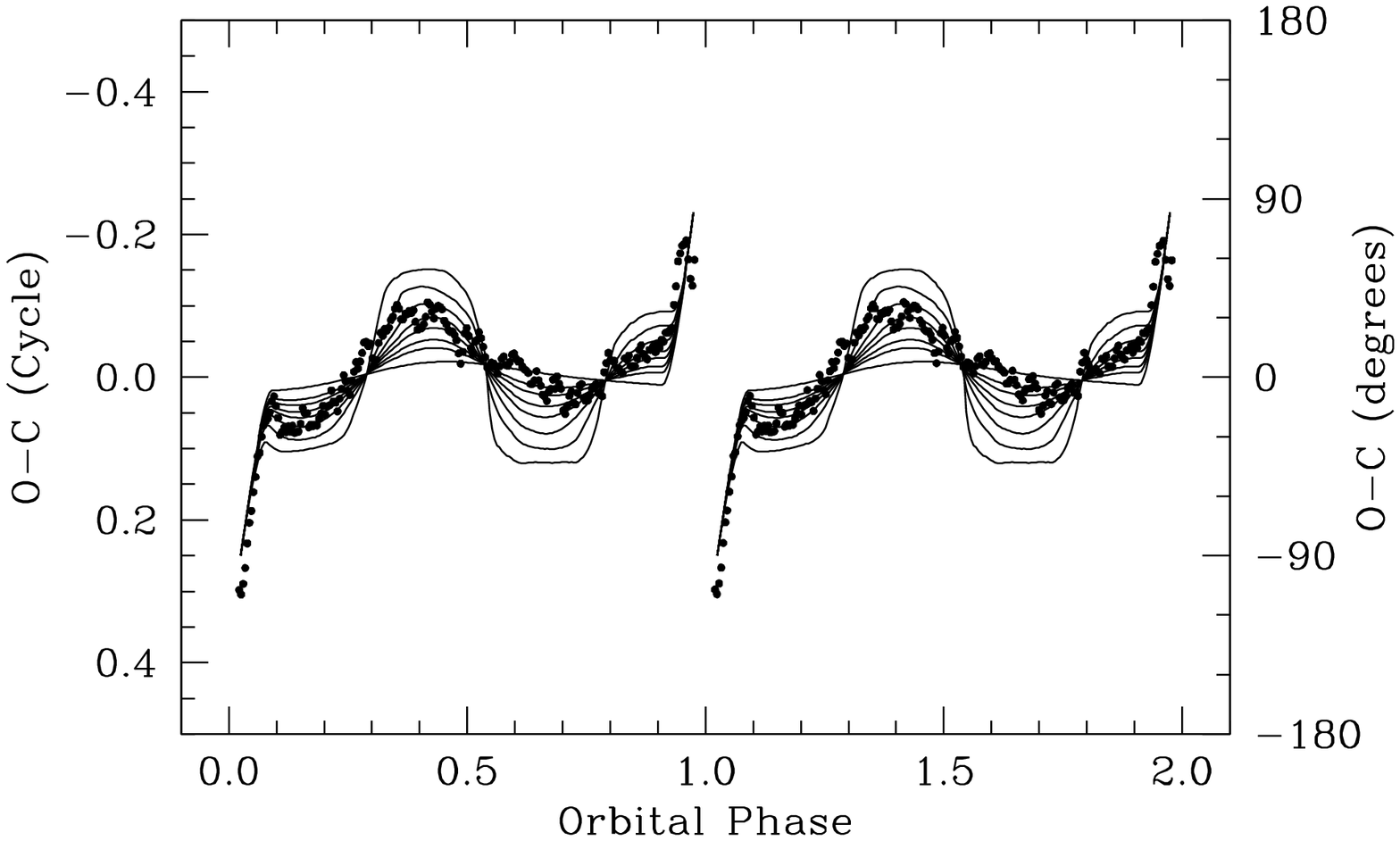,rwidth=10cm,width=12truecm,rheight=6.3cm,angle=0}}
{The $O-C$ diagram calculated from the SPH disc structure as a
function of orbital phase and for inclination angles $85^\circ$,
$89^\circ$, $89.3^\circ$, and $89.5^\circ$ through $89.9^\circ$ in
steps of $0.1^\circ$.  The results are very sensitive to inclination
angle, and suggest $89.5^\circ\le i\le89.7$.  }

\vfill\eject
\section{4. Conclusions}

\noindent
We can draw a number of conclusions from the above results.  
\medskip

\def\itbul#1{\item{$\bullet$}{#1}}
\itbul{The remarkable match between the observed and calculated $O-C$ phase
diagrams argues that the method of smoothed particle hydrodynamics
capable of reproducing the equilibrium disc structure, and statistics
can be improved by averaging the results over time.
Because we worked the forward problem and varied both viscosity
coefficients and particle numbers in the simulations and arrived at
the same match with the observational results, we can feel relatively 
certain that the structure we find closely approximates that in the
disc of DQ
Herculis itself.}

\itbul{Non-circular streamlines in the outer disc are responsible for the
extent and vertical structure of the reprocessing region and rim disc
self-eclipse profile.  The radius of the reprocessing region is only
about one-half the disc radius, and results from an inflection in
disk thickness vs.\ radius relation.  While the reprocessing region is
completely eclipsed, the outer disc is not, and this may in part
help explain the existence of UV emission lines during mid eclipse,
at which time the UV continuum is absent (Silber et al.\ 1996a,b).
These observations have also led to the suggestion that a wind is
present in the system (Eracleous et al 1998).
}

\itbul{The range of inclination angles which provide the best match between
the observational and model result $O-C$ phase diagrams is $89.5^\circ
\le i \le 89.7^\circ$.  The model results are relatively insensitive
to choice of which profile is chosen of the range 50-th to 98-th
percentile.  The accretion stream begins to appear in the 99-th
percentile profile, and because it does not appear to contribute to 
the $O-C$ phase variations, we conclude it is optically thin.}

\itbul{The $O-C$ phase variations are sensitive to small changes
($\pm0.1^\circ$) in disc inclination, and hence also to comparably
small changes in the disc opening angle.  Orbit-to-orbit as well as
year-to-year variations in the $O-C$ diagram are thus naturally
explained as resulting from 1-5\% changes in the disc thickness
resulting from for example turbulent or magnetic stresses, or slow
changes in the mass transfer rate.}

\itbul{The total amplitude of the $O-C$ phase variations is less than
0.5 cycles, and so the 71-s rotation period is favored.  However, the
phase variations just before and after totality of the reprocessing
shows that our model is still missing some important details.}

\medskip

It would be useful to run an $N\sim10^6$ simulation of DQ Her to
calculate with reasonable statistics the orbit-to-orbit variations of
the $O-C$ phase variations, and to compare these with those present in
the  WET observations.  It would also be interesting to use the
velocity and relative temperature fields of the ``ray-traced surface
particles'' (see Simpson, Wood, \& Burke 1998) as a basis for
calculating the spectrum as a function of orbital phase. Both projects
should provide even tighter constraints on fundamental system
parameters for DQ Herculis.

\vskip3mm
\noindent
ACKNOWLEDGEMENTS. We thank all the telescope allocation committees for
their support of international telescope collaboratives.  This work
was supported in part by NASA grant NAG 5-3103 to Florida Institute of
Technology.  
The observations at BAO
were supported by the Chinese Natural Science Foundation through grant
No.\ 19673008.  Astronomy
at the Wise Observatory is supported by grants of the Israel Science
Foundation.  G.V., N.D. and M.C. acknowledge telescope time allocation
at the Bernard Lyot Telescope of Pic du Midi Observatory, which is
operated by the CNRS.


\References 
\def\apj{ApJ}
\def\aj{AJ}
\def\araa{ARA\&A}

\def\mnras{MNRAS}
\def\pasp{PASP}

\ref
Bath, G. T., Evans, W. D., \& Pringle, J. E. 1974, MNRAS, 166, 113 

\ref
Chester, T.J. 1979, \apj, 230, 167

\ref
Eracleous, M., Livio, M., Williams, R. E., Horne, K., Patterson, J.,
Martell, P., \& Korista, K. T. 1998, in {ASP Conf Series \#137: Wild
Stars in the Old West}, Eds. S. Howell, E. Kuulkers, \& C. Woodward
(San Francisco: ASP), 438

\ref
Horne, K., Welsh, W.F., \& Wade, R.A. 1993, \apj, 410, 357 (HWW)

\ref
Kepler, S.O. 1990, Baltic Astron., 2, 515

\ref
Lamb, D. Q. 1974, ApJ, 192, L129

\ref
Lattanzio, J. C., Monaghan, J. J., Pongracic, H., \& Schwarz, M. P.
1986, J. Sci. Stat. Comput., 7, 591


\ref
Monaghan, J. J.  1992, \araa, 30, 543

\ref
O'Donoghue, D. 1985, in {Proc. 9\up{th} North Am. Workshop on
Cataclysmic Variable Stars}, ed. P. Szkody (Seattle: U. Washington) p.
98

\ref
Osaki, Y., Hirose, M. \& Ichikawa, S. 1993 in {Accretion Disks in
Compact Stellar Systems}, ed. J.C Wheeler (Singapore: World Sci.
Publ.) p.  272

\ref
Paczy\'nski, B.  1977, ApJ, 206, 822 

\ref
Patterson, J., Robinson, E.R., \& Nather, R.E. 1978,
\apj, 224, 570

\ref
Patterson, J. 1994, \pasp, 106, 209.

\ref
Petterson, J. 1980 \apj, 241, 247

\ref
Silber, A.D., Anderson, S.F., Margon, B., \& Downes, R.A. 1996,
\apj, 462, 428

\ref
Silber, A.D., Anderson, S.F., Margon, B., \& Downes, R.A. 1996b, \aj,
112, 1174

\ref
Simpson, J. C. 1995, \apj, 448, 822

\ref
Simpson, J. C., \&  Wood, M. A. 1998, \apj, 506, 360

\ref
Simpson, J.C., Wood, M. A., \& Burke, C.J.
        1998, {Baltic Astronomy}, {7}, 255

\ref
Walker, M.F. 1954, \pasp, 66, 230

\ref
Walker, M.F. 1956, \apj, 123, 68
\ref
Warner, B. 1995, Cataclysmic Variable Stars (Cambridge: Cambridge)

\ref
Warner, B., Peters, W. L., Hubbard, W. B., \& Nather, R.E. 1972,
        \mnras, 159, 321

\ref
Wood, M. A., \&  Simpson, J. C. 
1995, {Baltic Astronomy}, {\bf 4}, 402

\ref
Zhang, E., Robinson, E.R., Stiening, R.F., \& Horne, Keith 1995,
        \apj, 454, 447

}
\bye